\documentstyle[aps,preprint,epsf]{revtex}

\newcommand{\be}{\begin{eqnarray}}
\newcommand{\ee}{\end{eqnarray}}

\newcommand{\la}{\langle}
\newcommand{\ra}{\rangle}

\newcommand{\half}{ {\textstyle\frac{1}{2}} }
\renewcommand{\|}{ {{\scriptscriptstyle{\parallel}}} }
\renewcommand{\i}{ {\rm i} }
\renewcommand{\o}{\omega}
\renewcommand{\t}{\tilde}
\renewcommand{\vec}{\bbox}

\begin{document}

\draft

\title{Resolving the space-time structure of sonoluminescence by
  intensity interferometry}

\author{Claus Slotta and Ulrich Heinz}

\address{Institut f\"ur Theoretische Physik, Universit\"at Regensburg,
         D-93040 Regensburg, Germany}

\date{\today}

\maketitle

\begin{abstract}

We analyze the spatial and temporal resolving power of two-photon
intensity interferometry for the light emitting source in single
bubble sonoluminescence (SBSL). We show that bubble sizes between
several 10 nm and 3 $\mu$m can be resolved by measuring the transverse
correlation function, but that a direct determination of the flash
duration via the longitudinal correlation function works only for
SBSL pulses which are shorter than 0.1 ps. Larger pulse lengths can be
determined indirectly from the intercept of the angular correlator
at equal photon frequencies. The dynamics of the bubble is not
accessible by two-photon interferometry. 

\end{abstract}


\narrowtext


\section{Introduction}
\label{sec1}

In single bubble sonoluminescence (SBSL) \cite{report} an air
bubble in water is trapped in the velocity node of an acoustical
pressure field. Under defined conditions this bubble periodically
emits intense, broad-banded flashes of light, synchronous to the
driving sound. Measurements of pulse widths have given estimated
values from less than $50\,$ps \cite{report,moran} to more than
$250\,$ps \cite{Gompf97}, and an ana\-ly\-sis of the bubble surface's
temporal variation indicates minimal radii of $\sim 0.5\,\mu$m
\cite{report}. A direct measurement of the size of the light emitting
region does not exist so far.

The fundamental light generating mechanism is still unknown.
Some models \cite{Wu94,Hiller92} attribute the light emission to
electronic excitations or Bremsstrahlung processes initiated by
spherically converging shock fronts, leading essentially to a
black-body spectrum. While reproducing the measured spectra quite well
they require extraordinarily high temperatures with at least partial
ionization of the trapped gas. The peak of this thermal spectrum and
its high-energy Boltzmann tail are postulated to be hidden below
the absorption edge of water ($\lambda < 180\,$nm) where the photon
spectrum cannot be measured. Alternative explanations, invoking
collision-induced emission \cite{Frommhold94} or quantum vacuum
radiation \cite{Eberlein96}, operate at more moderate temperatures
and, in the latter case, do not exhibit an (invisible) high energy
component while still reproducing the spectral shape in the
measurable low energy window. Numerical simulations \cite{Wu94}
combining the gas dynamics inside a sphere with the oscillations of
the bubble surface yield pulse widths of $\sim 1\,$ps combined with a
radius of $\sim 0.5\,\mu$m for the emission region. These simulations
indeed produce strong shock waves which are seen as causal for the
light emission, but the predicted flash duration does not seem to
coincide with the experimental data.

Intensity interferometry, based on Bose-Einstein correlations (BEC) 
between identical bosons, has been discussed as a possibility to 
clarify the structure and dynamics of the light emitting region
\cite{Trentalange96,Hama97} by directly measuring its size and
lifetime. This technique, originally developed to determine the
angular size of stars \cite{Hanbury54}, has recently been considerably
refined in order to extract the spatial and temporal structure of the
hot reaction zone created in high energy nuclear collisions
\cite{Heinz97}. It has proven to be a valuable tool also in the 
analysis of {\em dynamical} processes in extremely small and short-lived
particle sources. Its application to SBSL in the form of two-photon
intensity interferometry should be further facilitated by the
comparatively large number of photons emitted per pulse and by the 
absence of many of the complications present in high energy
applications, like final state Coulomb and strong interactions among
the produced particles and secondary production of particles outside
the source by decay of unstable resonances. 

If successful, SBSL interferometry may in fact turn out to be a
unique testing ground for the method itself since here, contrary to
high energy physics where the sources are too small and shortlived to
be probed externally, alternative investigation techniques are
available which should allow for various cross checks.

In this paper we supplement the suggestion of 
\cite{Trentalange96,Hama97} with a general and largely model 
independent discussion of the method and its application to SBSL,
providing a {\em quantitative} analysis of its spatial and temporal
resolving power in the limited range of experimentally accessible 
wavelengths. We show that the transverse correlator is sensitive
to bubble sizes in the physically interesting domain while direct
resolution of the pulse length via the longitudinal correlator is
probably not possible with present technology. It can be determined
indirectly, however, via the intercept of the transverse correlator at
zero opening angle between the detectors. The dynamics of the bubble
can, unfortunately, not be resolved interferometrically.

\section{Two-photon correlation function} 
\label{sec2}

The correlation function for two photons with momenta
$\vec{k}_a$ and $\vec{k}_b$ is defined as
 \be
 \label{corrdef}
    C(\vec{k}_a,\vec{k}_b) = \frac{P_2(\vec{k}_a,\vec{k}_b)}
                             {P_1(\vec{k}_a)\,P_1(\vec{k}_b)} \;,
 \ee      
where $P_1(\vec{k})$ is the inclusive single-photon spectrum and
$P_2(\vec{k}_a,\vec{k}_b)$ is the two-photon coincidence spectrum. 
All photon energies are on-shell, $\o_{a,b} = |\vec{k}_{a,b}|$.
In the following we assume that the photons are emitted completely
incoherently and that the source is spherically symmetric. While the
latter is strongly suggested by the extreme stability of the
oscillating bubble, some phase coherence among the emitted photons can
presently not be excluded. Ideally one would test this by measuring
the (true) intercept of the two-photon correlator (\ref{corrdef}) at
$\vec{q} = \vec{k}_a - \vec{k}_b = 0$ \cite{Hama97}. As we will show
such a measurement is difficult, due to the limited frequency
resolution of present photon detectors; in the long run one might
therefore contemplate a comparison of 2- and 3-photon interferometry
data to settle the issue of partial coherence \cite{Zhang97}. 

For chaotic (incoherent) sources the single-photon spectrum
$P_1(\vec{k})$ and the correlator $C(\vec{k}_a,\vec{k}_b)$ can be
expressed in terms of the single-photon Wigner phase-space density
$S(x;K)$ of the emitting source \cite{Shuryak73,Pratt90,Chapman94}:
 \be
 \label{spec}
   P_1(\vec{k}) &=& \int d^4x\, S(x;\vec{k},\o)\, ,
 \\
   C(\vec{k}_a,\vec{k}_b) &=& 1+\frac{1}{2} 
  \frac{\left|\int d^4x\,S(x;K)\,e^{\i q\cdot x}\right|^2}
       {\int d^4x\,S(x;\vec{k}_a,\o_a)\,\int d^4y\,S(y;\vec{k}_b,\o_b)}
 \nonumber\\
 \label{corrapp}
   &\approx& 1+\frac{1}{2} 
  \left|\frac{\int d^4x\,S(x;\vec{K},E)\,e^{\i q\cdot x}}
             {\int d^4x\,S(x;\vec{K},E)}\right|^2 \, .
 \ee
Here $K=(\o_a+\o_b,\vec{k}_a+\vec{k}_b)/2$ and $q=(\o_a-\o_b, 
\vec{k}_a-\vec{k}_b)$. The second equation in (\ref{corrapp}) is an
approximation in that the single-photon spectra in the denominator
have been evaluated at the average momentum $\vec{K}$ rather than at
$\vec{k}_a$ and $\vec{k}_b$, and in both the numerator and denominator
the correct energy variables ($\o_a$, $\o_b$, and $K_0=(\o_a+\o_b)/2$,
respectively) have been replaced by the on-shell energy corresponding
to $\vec{K}$, $E=\vert\vec{k}_a + \vec{k}_b\vert/2$. This
approximation makes the following discussion more transparent, but can
be systematically corrected for \cite{Chapman95} (see below).
The factor ${1\over 2}$ in front of the second term in the
correlator (\ref{corrapp}) takes into account \cite{Slotta97} that
only photons with equal helicity states are affected by Bose
symmetrization. 

Since the measured photons are on-shell and thus the Fourier transform
in (\ref{corrapp}) is not invertible, the space-time structure of
$S(x;K)$ cannot be uniquely reconstructed. Still, valuable information
on the space-time structure of the source can be extracted from the
measured correlation function in terms of the second central
space-time moments of $S(x;K)$ \cite{Chapman95,Chapman95b,Heinz97}. In
the context of SBSL this will be discussed next.

\section{Geometrical analysis}
\label{sec3}

Detailed investigations (for a recent overview see
\cite{Heinz97}) have shown that the essential features of the correlator
(\ref{corrapp}) can be captured by replacing the $x$-dependence of
the emission function $S(x;K)$ by a Gaussian with the same center and
width. This is even more true for SBSL applications than in high energy
particle physics since here resonance decay effects which can invalidate
this Gaussian approximation are absent. Inserting a Gaussian ansatz
for $S(x;K)$ into (\ref{corrapp}) yields a correlator which is
Gaussian in the relative momentum $q$. Due to the spherical symmetry
of the problem there is only one distinguished direction which is
defined by the photon pair momentum $\vec{K}$. We therefore use a
Cartesian coordinate system in which $\vec{K} = (E,0,0)$ (i.e. all
$K$-dependence can be expressed through the energy $E$) and $\vec{q} =
(q_\|,q_\bot,0)$. We also have $q_0 = \o_a-\o_b = {E\over \o_a+\o_b}
q_\| \approx q_\|$. Following the techniques developed in
\cite{Chapman95,Chapman95b} it is then easily seen that the most
general form of the correlator reads
 \be
 \label{taylor}
   C(\vec{q},E) \approx 1 + \half  \, 
   e^{- q_\bot^2\,\left\la x_\bot^2\right\ra(E)  
      - q_\|^2\,\left\la (\t{x}_\| - c \t{t})^2 \right\ra(E)} \, ,
 \ee
where the angular brackets denote averages taken with the source
function,
 \be
 \label{variance}
   \la f(x) \ra(E) = \frac {\int d^4x\, f(x)\, S(x;E)}
                           {\int d^4x\, S(x;E)} \;,
 \ee
and tilde superscripts indicate center-corrected coordinates, $\t{x}_i
= x_i - \la x_i \ra(E)$. (Note that $\langle x_\bot \rangle(E)=0$ due
to spherical symmetry.) Eq.~(\ref{taylor}) tells us that by measuring,
at fixed $E$, the correlator as a function of $q_\bot$ and $q_\|$,
respectively, we can determine the spatial variance $\la x_\bot^2\ra$ 
and the mixed variance $\la (c\t{t} - \t{x}_\|)^2 \ra$,
respectively, of the effective source of photons with energy $E$.
In principle, for different $E$ the effective source can have
different such ``sizes'' or ``HBT radii''.

In relativistic heavy ion collisions the $K$-dependence (here:
$E$-dependence) of the space-time variances plays an important role as
a signature for collective expansion of the emitting source
\cite{Heinz97}. This is most easi\-ly seen in the context of a
hydrodynamically expanding, locally thermalized source whose momentum
dependence is dominated by a boosted Boltzmann distribution $\sim \exp
[-K\cdot u(x)/T(x)]$ where $u(x)$ is the collective expansion 4-velocity
profile. This factor generates correlations between the momentum $K$
and the position $x$ in the emitter which in turn cause a
$K$-dependence of the HBT radii. The strength of these correlations
can be estimated by writing
 \be
 \label{estimate}
   S(x;E) \sim s(K{\cdot}u(r,t)/T) \approx s(E/T) 
   \bigl(1 + {\cal O}(v/c)\bigr) \;.
 \ee
Since the expansion velocity $v(r,t)$ of the bubble in SBSL is limited
by the shock velocity in the compressed bubble gas and thus below
about $3\times 10^4$ m/s \cite{Wu94} (i.e. $v/c < 10^{-4}$), the
nonrelativistic estimate (\ref{estimate}) is reliable and the $x$-$K$
correlations induced by the collective dynamics of the bubble are
weak. This is different for pion interferometry in heavy ion
collisions: there the collective velocities are of the order of the
light velocity, causing strong $x$-$K$ correlations and an appreciable
$K$-dependence of the correlator which can be used as a dia\-gnostic
tool \cite{Heinz97}. In SBSL interferometry, on the other hand, the
$E$-dependence of the HBT radii resulting from the weak $x$-$K$
correlations is so small that it can be neglected in the measurable
$E$-range (see below). This is unfortunate since it means that SBSL
interferometry will not give any direct access to the collective
dynamics of the bubble during light emission \cite{fn1}. On the other
hand, it simplifies the theoretical description because we can neglect
the $E$-dependence of the HBT radii and also drop the cross term in
the longitudinal HBT radius: 
 \be
 \label{nocross}
   \left\la (\t{x}_\| - c \t{t})^2 \right\ra \approx 
   \left\la x_\|^2 \right\ra + c^2 \left\la \t{t}^2 \right\ra \, .
 \ee    
This is true because both the displacement $\la x_\| \ra$ of the
source center in $\bbox{K}$-direction and the cross term $\la x_\|\,
t\ra$ are also generated by the collective expansion \cite{Chapman95}
and thus here expected to be about 4 orders of magnitude smaller than
the diagonal terms $\la x_\|^2 \ra$ and $\la \t{t}^2 \ra$. 

The corrections resulting from the approximation (\ref{corrapp}) can
now be systematically included following the discussion in
\cite{Chapman95}. Writing up to second order in $q$
 \be
   {(\o_a+\o_b)^2 \over 4} 
   \approx E^2 + \frac{1}{4} \left( \vec{q}^2 -
                       \frac{(\vec{K}\cdot\vec{q})^2}{E^2}\right) \; ,
 \ee
one derives to quadratic accuracy 
 \be
 \nonumber
   \int d^4x\,S(x;\vec{K},K_0) \: e^{\i q\cdot x} \approx 
   e^{ \Delta R_\bot^2  q_\bot^2} 
   \int d^4x\,S(x;\vec{K}) \: e^{\i q\cdot x}
 \ee
with
 \be
   \Delta R_\bot^2(E) = \left( \frac{1}{8 E} 
   \frac{\rm d}{{\rm d} E} \ln P_1(E)\right)\;.
 \ee
The single particle distribution $P_1(\o_a)$ may be similarly
approximated by 
 \begin{eqnarray}
   P_1(\o_a) \approx
      \left( 1+ \left(\frac{q_\bot^2}{8 E} + \frac{q_\|}{2}\right) 
             \frac{\rm d}{{\rm d} E} + \frac{q_\|^2}{8} 
             \frac{{\rm d}^2}{{\rm d} E^2}  
             \right) P_1(E)  .
 \end{eqnarray}
For $P_1(\o_b)$ one obtains the same expression with the opposite sign
for the term linear in $q_\|$. For the denominator in (\ref{corrapp})
one thus finds to quadratic order 
 \be
   P_1(\o_a) P_1(\o_b) \approx  
   e^{ 2\Delta R_\bot^2 q_\bot^2 + \Delta R_\|^2 q_\|^2 } \: 
   P_1^2(E)
 \ee
with
 \be 
   \Delta R_\|^2(E) = \frac{1}{4}\frac{{\rm d}^2}{{\rm d} E^2} \ln P_1(E)\;.
 \ee
Hence the corrected Gaussian expression (\ref{taylor}) for the correlator
reads  
 \begin{mathletters}
 \label{summary}
 \be
 \label{correl}
   C(q_\bot,q_\|,E) &\approx& 1 
   + \half \: e^{-R_\bot^2\:q_\bot^2 - R_\|^2 \: q_\|^2} \, ,
 \\
 \label{Rperp}
   R_\bot^2 &=& \la x_\bot^2\ra + \Delta R_\bot^2 \, ,
 \\ 
 \label{Rpar}
   R_\|^2   &=& \la   {x}_\|^2 \ra + c^2 \la \t{t}^2 \ra  
              + \Delta R_\|^2  \, .
 \ee
 \end{mathletters}
Note that the corrections $\Delta R_\bot^2$ and $\Delta R_\|^2$
are proportional to the slope and curvature of the logarithmic
intensity spectrum, respectively, and are therefore directly 
accessible from single-photon measurements. Both $\Delta R_\bot$ and
$\Delta R_\|$ turn out to be at most several 10~nm. 

The experimental realization of the correlation measurement (as
proposed by Trentalange and Pandey \cite{Trentalange96}) consists of
two photo-multipliers focussing on the sonoluminating bubble at a
relative angle $\phi$. The required momentum resolution is achieved
by suitably chosen apertures and pre-detector band-pass filters. The
signal detected in one multiplier during one flash, proportional to
the incident number of photons, is correlated with the output of the
second device during the same flash and sampled over a sufficient number 
of bubble oscillations for statistics.

This experimental setup suggests the use of $q_0 = \o_a-\o_b$
and $\phi$ instead of the variables $q_\perp$ and $q_\|$. They are
related via
 \begin{mathletters}
 \label{qrel}
 \begin{eqnarray}
 \label{qrel1}
   q_\bot^2 &=& \left( 4E^2 + \frac{q_0^4}{4E^2} - 2q_0^2 \right)  
                \tan^2 \half\phi \;,
 \\
 \label{qrel2}
   q_\|^2 &=& q_0^2 + \left( q_0^2 - \frac{q_0^4}{4E^2} \right)
                         \tan^2 \half \phi  \;.
 \end{eqnarray}
 \end{mathletters}
$R_\perp$ and $R_\|$ can thus be isolated by fixing the average photon
energy $E$ and scanning the correlator either as a function of the
opening angle at equal photon energies ($q_0=0$, ``transverse
correlator''), or as a function of the energy difference $q_0$ at zero
opening angle ($\phi=0$, ``longitudinal correlator''):
 \begin{mathletters}
 \label{fit}
 \begin{eqnarray}
 \label{fit1}
   C(q_0=0,\phi,E) &\approx& 1 
   + \half\: \exp \left(- R_\bot^2 \: 4E^2 \tan^2 \half\phi\right) \, ,
 \\ 
 \label{fit2}
   C(q_0,\phi=0,E) &\approx& 1 
   + \half\: \exp \left(- R_\|^2 \: q_0^2 \right) \;.
 \end{eqnarray}
 \end{mathletters}

\section{Resolving power}
\label{sec4}

We will now proceed towards a quantitative estimate of the
resolving power of such correlation measurements. We begin by noting
that our approximations break down if the first terms on the right
hand sides of (\ref{summary}b,c) become smaller than the
corrections from the second terms. This turns out not to be the
limiting factor, though, because similar lower limits for the HBT
radii result from the fact that the opacity of water prohibits
measurements at wavelengths in the ultraviolet below 180 nm, and low
yields make the measurement difficult in the infrared ($\o \alt
1.5$ eV). Practical measurements are only possible in the
``transparency window'' 1.5~eV $\alt \o \alt$ 6~eV (210~nm
$\alt \lambda \alt$ 830~nm). As we will see this limits, at
fixed $E$, the opening angle $\phi$ and the energy difference $q_0$ 
which means that the correlator (\ref{fit}) can only be measured 
over a restricted interval of the control variables. 
If the correlator does not fall off appreciably over the accessible 
range, the HBT radius parameter cannot be accurately determined. 
This gives lower limits for $R_\perp$ and $R_\|$ of several 10~nm, 
i.e. of the same order of magnitude as the upper limits for the 
correction terms $\Delta R_\|,\Delta R_\perp$.

\begin{figure}\epsfxsize=6cm
\centerline{\epsfbox{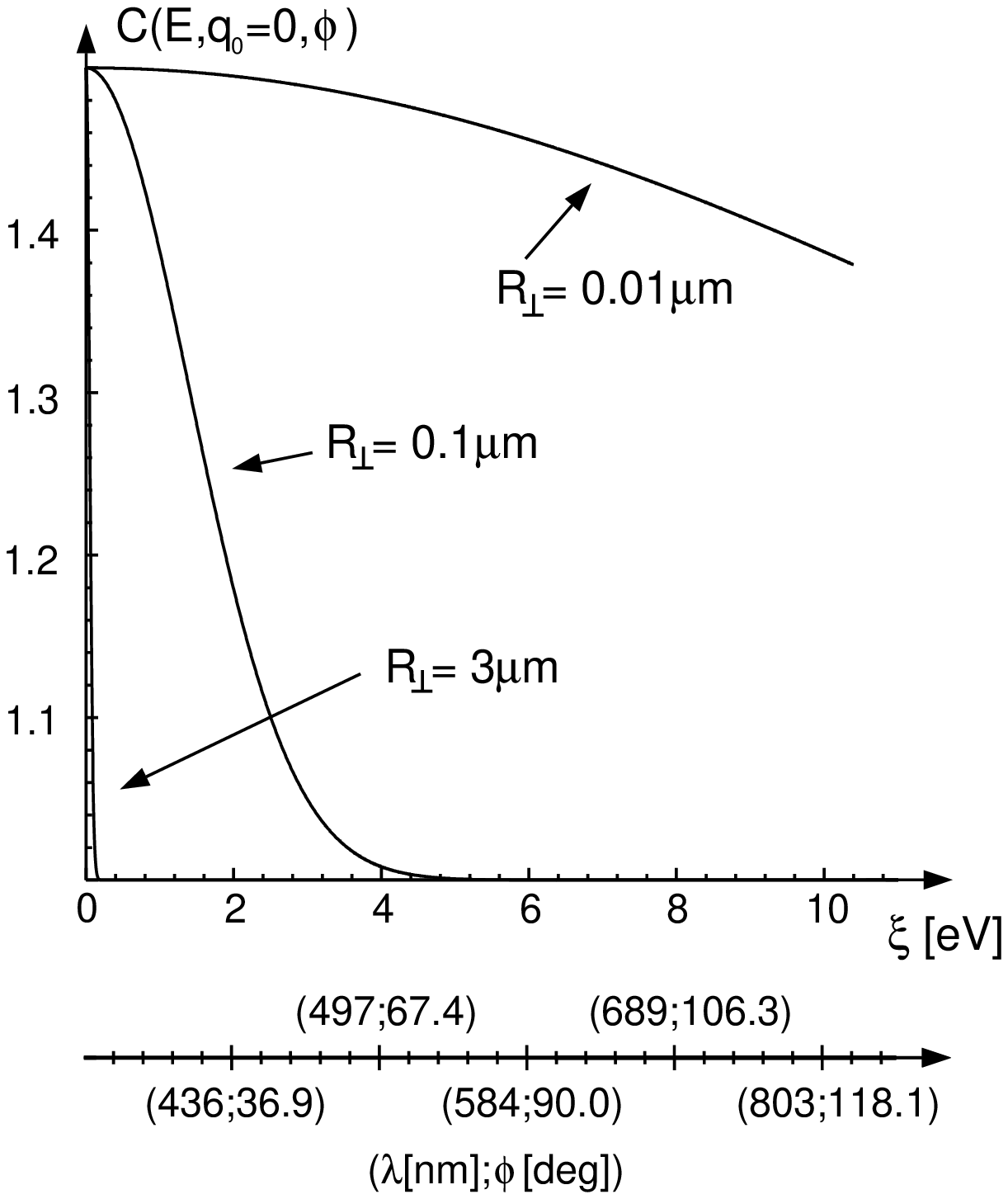}}
\vskip4mm
\caption{The correlation function $C(E=3\,{\rm eV},q_0 = 0,\phi)$
  as a function of $\xi = 2 E \tan\half\phi$ for various values
  of $R_\bot$. The domain above $\xi \approx 10.4\,$eV is not
  accessible due to light absorption in water. The second abscissa 
  gives the detector settings for some typical $\xi$ values.}
\label{fig1}
\end{figure}

After these general remarks let us enter a more detailed discussion,
beginning with $R_\perp$. In Fig.~\ref{fig1} we have plotted the
correlator (\ref{fit1}) at fixed $E$ = 3~eV and $q_0=0$ as a function
of the opening angle $\phi$. One sees that for $R_\perp=10$~nm the
correlator falls off only by about 20\% over the measurable
range, rendering the determination of $R_\perp$ difficult. According to
Fig.~\ref{fig1}, good measurements of $R_\perp$ are possible for 10~nm
$< R_\perp <$ 3 $\mu$m. For $R_\perp > 3$ $\mu$m angular resolution
becomes a problem: the correlator falls off so rapidly that opening
angles between the two detectors and angular apertures of each
detector below $1^\circ$ are required to resolve the correlation
function. This obviously cuts down on event statistics. However,
$R_\perp > 3\,\mu$m implies a source with transverse size $\sqrt{\la
  x_\perp^2   \ra} \agt 3\, \mu$m, see Eq.~(\ref{Rperp}). Since it
is known that at the point of light emission the source is smaller
than this (values below 1 $\mu$m are quoted in \cite{report}), angular
resolution of the measurement does not appear to be a crucial limiting
factor.  
 

One should note, however, that $q_0=0$ as indicated in Fig.~\ref{fig1}
implies ideal energy resolution of the photon detector. We will see
shortly that the finite energy resolution in real life modifies
significantly the optimistic picture suggested by Fig.~\ref{fig1}.

Let us now turn to a discussion of $R_\|$. It is easy to see that if
the light emitting source has a radius below 1~$\mu$m
(i.e. $\sqrt{\parbox[b]{1.8em}{\rule[-0.5ex]{0mm}{2.5ex}}}  
\hspace*{-1.8em}{\la{x}_\|^2 \ra}   
\alt 1\,\mu$m) then the r.h.s. of
(\ref{Rpar}) is dominated by the duration of the light flash
$\delta\tau = \sqrt{\la\t{t}^2\ra}$ as soon as $\delta\tau$ becomes
larger than about 3 femtoseconds (which corresponds to $c\,\delta\tau
= 1\, \mu$m). Since typical SBSL pulse durations discussed in the
literature \cite{report,moran,Gompf97} are much longer we can for the
following discussion neglect in (\ref{Rpar}) the geometric
contribution as well as $\Delta R_\perp$ and write $R_\| \approx c\,
\delta\tau$. 

\begin{figure}\epsfxsize=6.25cm
\centerline{\epsfbox{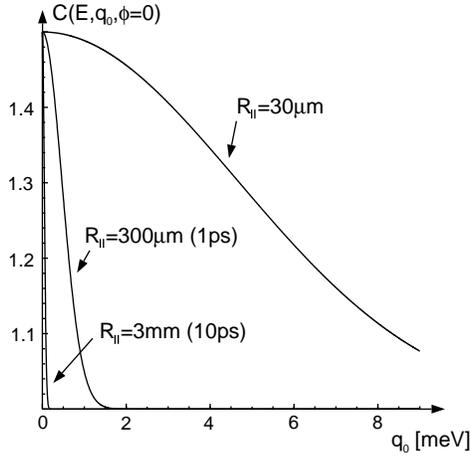}}
\vskip4mm
\caption{The correlation function $C(E=3\,{\rm eV},q_0,\phi=0)$
   as a function of the energy difference $q_0$, for various 
   values of $R_\perp$. One sees that pulse lengths above 1 ps 
   require an energy resolution well below 1 meV.}
\label{fig2}
\end{figure}

Fig.~\ref{fig2} shows that SBSL pulses which last longer than 1 ps
can only be resolved if the photon detector has an energy resolution
well below 1 meV (!). This implies a relative band width
$\delta\lambda / \lambda \alt 10^{-4}$. As we will see in a moment,
it is not sufficient that the filter-to-filter distance between the
two detectors is known with this accuracy; the band width of {\em each
filter individually} must satisfy this constraint.

Commercially available filters in the visible region around 400 nm
have band widths $\delta\lambda \agt 1$~nm, corresponding to
$\delta\lambda / \lambda \agt 25 \times 10^{-4}$. For
$\delta\lambda = 10$~nm the authors of \cite{Trentalange96} quoted
coincidence rates of 200-300 counts/s at a bubble-detector distance of
200 mm. For smaller $\delta\lambda$ the coincidence rate drops
essentially like $(\delta\lambda)^2$. 

Too large values of $\delta\lambda/\lambda$ imply that the correlator
in Fig.~\ref{fig2} is averaged over a range $q_0$ which is much
larger than the region over which the correlator drops back to 1. This
implies not only that the longitudinal correlation function
$C(q_0)$ cannot be resolved, but also that the transverse
correlation function $C(\phi)$, being averaged over a wide
$q_0$-range, is strongly diluted. As a consequence $C(\phi)$ will not
intercept the vertical $(\phi=0)$-axis at the ideal value
$1+{1\over 2}={3\over 2}$, but at a much lower value. The {\em 
effective intercept} will be the smaller the larger the band width 
of the photon detector. 

It is not difficult to calculate the effective intercept value as a
function of the ratio between the filter band width and the flash
duration. Let us assume filters with a Gaussian frequency profile
 \be
 \label{filter}
    f_{\bar\o,\delta\o}(\o) = {1\over\sqrt{2\pi(\delta\o)^2}}\,
    \exp\left[ - {(\o-\bar\o)^2 \over 2 (\delta\o)^2}\right]\, .
 \ee
One easily checks that
 \be
 \label{filter1}
    f_{\o_a,\delta\o}(\o_1)\, f_{\o_b,\delta\o}(\o_2) = 
    f_{K_0,\delta\o'}(\o)\, f_{q_0,2\delta\o'}(\Delta\o)
 \ee
where $\delta\o' = \delta\o/\sqrt{2}$, $\o = (\o_1+\o_2)/2$, $\Delta\o
= \o_1-\o_2$, and $K_0=(\o_a+\o_b)/2$, $q_0 = \o_a-\o_b$ as before.
Neglecting $x$-$K$ correlations in the source as discussed above we
can assume that $S(x;K)$ factorizes, $S(x;E) \approx X(x)\cdot s(E)$,
where $X(x)$ is normalized, $\int d^4x\, X(x) = 1$. For the pro\-duct of
single particle spectra in the denominator of Eq.~(\ref{corrapp}) we
thus obtain
 \be
 \label{denom}
 \FL
   &&P_1(\bbox{k}_a)\, P_1(\bbox{k}_b) =
 \nonumber\\
   &&\int d\o_1\, d\o_2\, f_{\o_a,\delta\o}(\o_1) 
     \,f_{\o_b,\delta\o}(\o_2) \, s(\o_1)\, s(\o_2) =
 \nonumber\\
   &&\int d\o\, f_{K_0,\delta\o'}(\o) \, \int d(\Delta\o)\, 
     f_{q_0,2\delta\o'}(\Delta\o) \, 
 \nonumber\\
   &&\qquad\qquad \times\,
     s\left(\o+\half\Delta\o\right)\, s\left(\o-\half\Delta\o\right) \, ,
 \ee
while the numerator is similarly found to be
 \be
 \label{num}
 \FL 
  &&P_2(\bbox{k}_a,\bbox{k}_b) - P_1(\bbox{k}_a)\, P_1(\bbox{k}_b) = 
 \nonumber\\
  &&{1\over 2} \int d\o \, \left(s(\o)\right)^2 \, f_{K_0,\delta\o'}(\o) 
    \int d^4x\, d^4y\, X(x)\, X(y)\, 
 \nonumber\\
  &&\times \, e^{-i\o(\bbox{x}-\bbox{y}){\cdot}(\bbox{e}_a-\bbox{e}_b)}   
           \int d(\Delta\o)\, f_{q_0,2\delta\o'}(\Delta\o) \, 
 \nonumber\\
  &&\times\, e^{i\Delta\o \left[(x^0-y^0) 
    - {1\over 2}(\bbox{x}-\bbox{y}){\cdot}(\bbox{e}_a+\bbox{e}_b)\right]}
    \, ,
 \ee
where $\vec{e}_{a,b}$ are unit vectors in direction of
$\vec{k}_{a,b}$. Since the filter band width $\delta\o$ is narrow, the
single particle spectrum $s(E)$ can be taken constant inside the
filter gap. This so-called ``smoothness approximation'' allows to
perform the integration over $\Delta\o$ in (\ref{denom}):
 \be
 \label{denom1}
   P_1(\bbox{k}_a)\, P_1(\bbox{k}_b) \approx 
   \int d\o \, \left(s(\o)\right)^2\, f_{K_0,\delta\o'}(\o)  \, .
 \ee
To obtain the effective intercept we divide (\ref{num}) by
(\ref{denom}) and set $k_a=k_b=K$, i.e. $q_0=\phi=0$ and $\vec{e}_a =
\vec{e}_b$. Then the $\o$-integration in Eq.~(\ref{num}) factorizes,
and the first factor on the r.h.s. of Eq.~(\ref{num}) cancels against
(\ref{denom1}). The second factor can be easily evaluated in the
Gaussian approximation where we replace the space-time factor $X(x)$
by a Gaussian with the same rms widths. We find 
 \be
 \label{intercept}
   C(q=0) &=& 1 + {1\over 2}\,{1\over 
                \sqrt{ 1 + 4 (\delta\o)^2 
   \left(\la {x}_\|^2 \ra + c^2 \la\t{t}^2\ra \right)/(\hbar c)^2}}  
 \nonumber\\
   &=& 1 + {1\over 2}\, {1\over 
       \sqrt{ 1 + 4 (\delta\o)^2 R_\|^2/(\hbar c)^2}} \, .
 \ee

\begin{figure}\epsfxsize=6.8cm
\centerline{\epsfbox{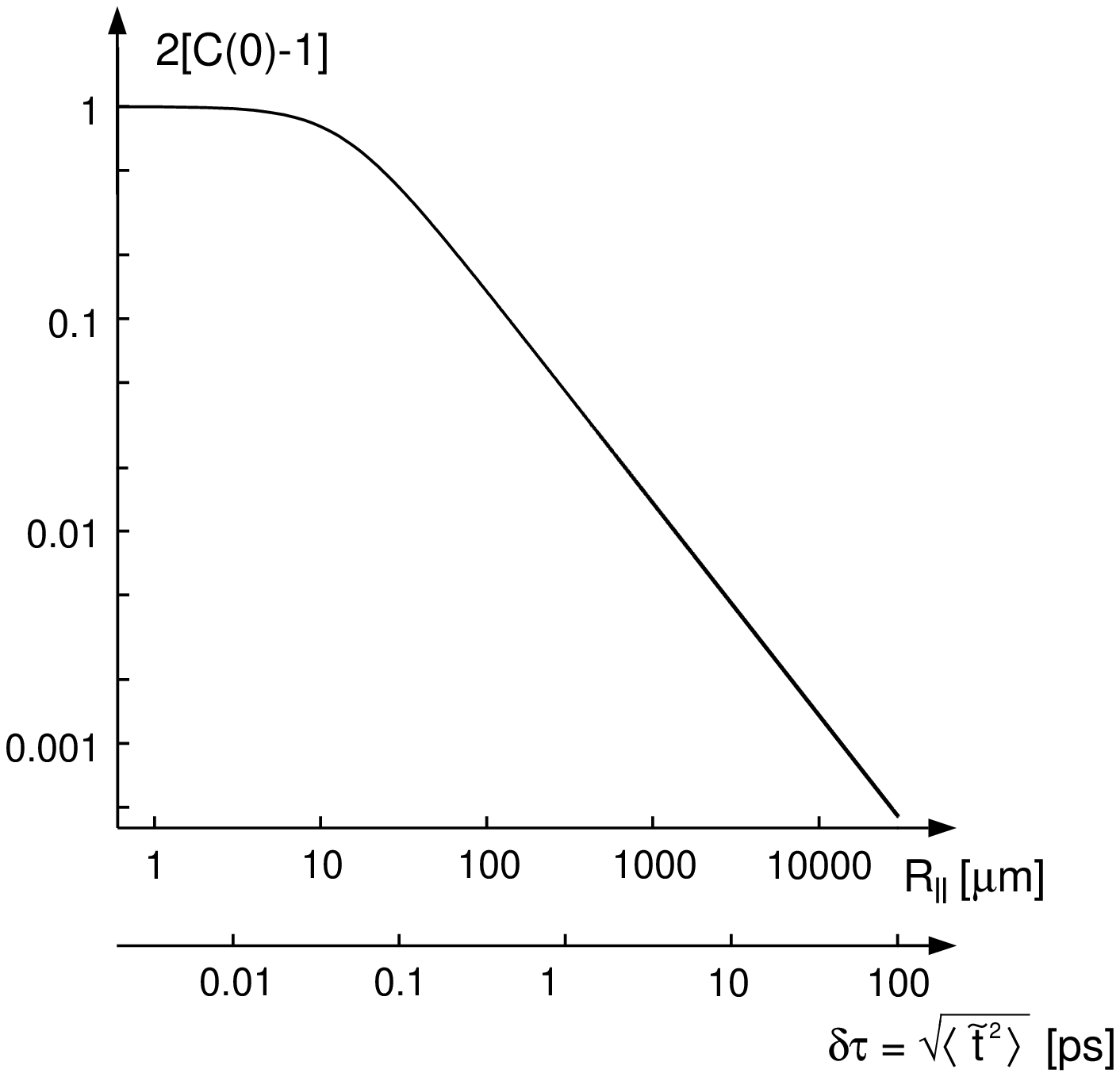}}
\vskip4mm
\caption{Effective intercept as a function of the flash duration 
     $R_\| \approx c \sqrt{\la \t{t}^2\ra}$ assuming a filter
     band width $\delta\lambda = 1\,$nm at $\lambda=413$~nm.}
\label{fig3}
\end{figure}

The dependence of this effective intercept on $R_\|$ (which in the
limit considered here is approximately equal to the flash duration
$c\,\delta\tau$) is plotted in Fig.~\ref{fig3} for a fixed filter band
width $\delta\lambda=1$~nm at an average photon ener\-gy $E=3$~eV 
(corresponding to $\lambda=413$~nm). For this case $\delta\o\,
R_\|/(\hbar c)=1$ corresponds to a flash duration $\delta\tau = 88\,{\rm
  fs} \approx 0.1\,$ps. One sees that the effective intercept is unity
for $\delta\tau \ll 0.1\,$ps; this is the domain where the given
filter band width allows to resolve the flash duration by measuring
the longitudinal correlation function (\ref{fit2}). For $\delta\tau
\gg 0.1\,$ps the effective intercept decreases linearly with the pulse
length, $C(0)-1 \sim 1/\delta\tau$. Thus, assuming completely
incoherent photon emission and $c\,\delta\tau \gg 
\sqrt{\parbox[b]{1.8em}{\rule[-0.5ex]{0mm}{2.5ex}}}  
\hspace*{-1.8em}{\la{x}_\|^2 \ra} $,  
$\delta\tau$ can be determined from the effective 
intercept of the transverse two-photon correlator (\ref{fit1}) even if
the longitudinal correlator (\ref{fit2}) does not show any structure
for $\delta$-values outside the experimental resolution (filter band
width) $\delta\o$. 


It is worth pointing out that according to this analysis it is not
necessarily advisable to strive for increasingly better filter
resolution $\delta\o$ resp. $\delta\lambda$. As discussed above,
increasing the band width enhances the coincidence rate 
quadratically while the effective intercept value decreases only
linearly. In leading order the effect of $\delta\lambda$ on the
experimental error bar of $\delta\tau$ thus cancels. Of course,
determining $\delta\tau$ via the effective intercept of the correlator
is a somewhat roundabout procedure which depends in a crucial way on
the assumed chaoticity of the source; eventually one would like to
achieve a genuine lifetime determination by measuring the longitudinal
correlator (\ref{fit2}) with appropriate frequency resolution. 

\section{Summary}
\label{sec5}

We have shown that two-photon intensity
interferometry can be applied to study the size and lifetime of the
light-emitting region in single-bubble sonoluminescence. A measurement
of the transverse correlation function in the experimentally accessible
frequency range provides sensitivity to sizes between several 
10 nm and a few $\mu$m for the active bubble region. Present
technological li\-mi\-tations on the frequency resolution limit a direct
measurement of the flash duration via the {\em longitudinal}
correlation function to pulse lengths below 0.1 ps. We showed,
however, that for chaotic emitters with longer flashes the pulse
duration can be determined indirectly via the intercept of the {\em
  transverse} correlation function as a function of frequency
resolution. The dynamics of the bubble during light emission is not
accessible by two-photon interferometry, due to the much too small
expansion velocities.   

\acknowledgments

U.H. would like to thank B.~Svetitsky for interesting discussions
which motivated this work. He also gratefully acknowledges clarifying
conversations with A.~Chodos, Y.~Ha\-ma, T.~Hemmick, G.~Kunde,
B.~Lasiuk, S.~Pa\-du\-la, and S.~Trentalange. This work was supported
by DFG, BMBF and GSI.  



\begin{references}
\bibitem{fn1}
   In the quantum vacuum radiation scenario of
   Ref.~\protect\cite{Eberlein96} time profiles $R(t)$ for the bubble
   radius were used which lead to much larger radial velocities which 
   might indeed be measurable via an $E$-dependence of the HBT
   radii. As pointed out in \protect\cite{Eberlein96} these large
   radial velocities are not really necessary for the scenario to work
   because what really matters for the light generation process is the
   4$^{\rm th}$ time derivative of the radial velocity at the
   turnaround point. Still, establishing the (approximate)
   $E$-independence of the HBT radii conjectured by us would put
   significant empirical constraints on dynamical scenarios like the
   one discussed in \protect\cite{Eberlein96}.
\end{references}
\end{document}